# Design, fabrication and characterization of the first AC-coupled silicon microstrip sensors in India

T. Aziz,[a] S.R. Chendvankar,[a] G.B. Mohanty,[a,*] M.R. Patil,[a] K.K. Rao,[a] Y.R. Rani,[b] Y.P.P. Rao,[b] H. Behnamian,[c,d] S. Mersi,[c] M. Naseri[c,d]

[a] *Tata Institute of Fundamental Research (TIFR), Mumbai, India*
[b] *Bharat Electronics Limited (BEL), Bangalore, India*
[c] *European Organization for Nuclear Research (CERN), Geneva, Switzerland*
[d] *Institute of Research in Fundamental Science (IPM), Tehran, Iran*

[*]*Tel:* +91-22-22782147, *Fax:* +91-22-22804610, *E-mail*: gmohanty@tifr.res.in

ABSTRACT: This paper reports the design, fabrication and characterization of single-sided silicon microstrip sensors with integrated biasing resistors and coupling capacitors, produced for the first time in India. We have first developed a prototype sensor on a four-inch wafer. After finding suitable test procedures for characterizing these AC coupled sensors, we fine-tuned various process parameters in order to produce sensors of the desired specifications.

# Contents

## 1. Introduction

Silicon microstrip sensors constitute an important component for various collider-based experiments at laboratories around the globe, e.g., CERN, Fermilab and KEK. These sensors are used because of their excellent spatial resolution, high detection efficiency and fast response time for tracking charged particles in multiparticle environments. Especially having a spatial resolution of a few tens of microns, they play a key role on locating secondary vertices of the short-lived particles that decay not-too-far-away from the primary interaction point.

     With the above motivations in mind, we have designed single-sided microstrip sensors (AC coupled) with integrated biasing resistors and coupling capacitors. They are fabricated at Bharat Electronics Limited [1] in Bangalore. Several batches of sensors have been studied and their process parameters optimized to achieve desired specifications. We have measured different characteristics for those sensors that came closest to our required specifications. We describe herein the design, fabrication, and various characterizations including a laser test and radioactive source measurement of the AC coupled sensors.

## 2. Silicon microstrip sensors

Silicon microstrip sensors are simply an array of p-n junction diodes, which are operated in a reverse bias mode such that full depletion occurs, clearing the bulk of any free charge carriers. When an external charged particle enters the fully depleted region of the sensor, it loses a part of its energy that is subsequently converted into electron-hole pairs along the path of the particle. The applied electric field prevents these charge carriers from recombining. Holes having positive charge move towards the negative electrode (p strips in the case of n-bulk material) while electrons travel to the positive electrode (backplane). The average energy loss per unit length is about 3.6 MeV/cm for a minimum ionizing particle (MIP). On the other hand, the energy required to create an electron-hole pair in silicon is 3.6 eV. Therefore, with a small band gap material such as silicon (band gap is 1.1 eV), a large number of electron-hole pairs are created for a given energy. One MIP creates approximately 24,000 electron-hole pairs in 300 μm silicon, resulting in a large signal output. A high mobility [2] value for electrons and holes (1350 and 480 cm$^2$/Vs, respectively) gives a fast signal.

### 2.1 Design of prototype sensors

Initially we have developed prototype sensors with a <111> crystal orientation and resistivities of 2 to 4 kΩ-cm. Silicon wafers of n type with a thickness of 300 μm and a diameter of 4 inch are used for this purpose. We have performed the design using the Cadence full development software [3]. The sensors are divided into 11 regions, each denoting a set, which have different strip width and pitch combinations. The overall sensor dimensions are 76 mm × 47 mm. Minimum (maximum) strip width and pitch of the sensors are 12 μm (48 μm) and 65 μm (135 μm), respectively. The number of strips per set varies between 30 and 32. Each p+ strip is connected to a common bias line through a polysilicon resistor. We have provided an independent bias pad for each set, AC and DC pads for each strip, and a polysilicon resistor between the bias pad and the DC pad. The polysilicon resistors are designed such that even strip resistors are located on one side and odd strip ones are on the other in order to provide sufficient space for them. The polysilicon resistor from each strip is connected to the common bias line, and each strip is AC coupled to the readout electronics to avoid saturation of the front-end amplifiers owing to large leakage currents on individual strips. Figure 1 shows a typical prototype sensor.

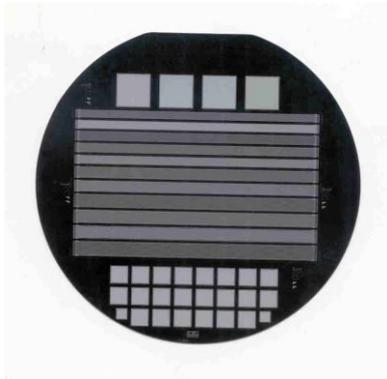

**Figure 1.** Photo of a prototype sensor.

The prototypes are a seven mask layer design. Individual masks are needed for the following processing steps:
1) p+ implant for each strip;
2) Silicon oxide for DC pads, bias line and p+ strips;
3) Polysilicon contact opening on DC pads;
4) Polysilicon between DC pads and bias line;
5) Contact opening for DC pads and bias line;
6) Metallization for AC pads, DC pads and bias line;
7) Protective layer for AC pads, DC pads and bias line.

### 2.2 Design of single-sided sensors

The overall dimension for single-sided sensors is 79.6 mm × 28.4 mm. The sensors are an eight mask layer design, similar to the prototype, but with an additional mask for deep n+

implantation. One guard ring is provided that encloses all the strips and has a width of 12 μm. AC pads for each strip are given on both ends of the strip in two rows along with the bias pad, as shown in Figure 2. The AC pad is appropriately dimensioned for wire bonding with the readout electronics. The polysilicon resistor setup is similar to the prototype sensor. For all the batches we have used the specifications listed in Table 1. A corner of the sensor showing the AC pads and the common bias line is shown in the left-hand photo of Figure 2, while the photo on the right shows the DC pads, polysilicon resistor structures and p+ strips.

**Table 1.** Specification of single-sided sensors.

| Wafer | n type silicon, 4 inch diameter, 300 μm thickness |
|---|---|
| Implantation for each strips | p+ |
| Number of strips | 1024 |
| Strip width | 50 μm |
| Strip pitch | 75 μm |
| Length of the strip | 25.6 mm |
| Active area of the sensor | 76.8 mm × 25.6 mm |
| Total leakage current | < 2 μA at 100 V (reverse voltage) |
| Polysilicon resistance | 4 MΩ ± 10% tolerance |
| Coupling capacitance | 150 pF ± 10% tolerance |
| Fraction of strips with pinholes | < 1% |
| Breakdown voltage | 300 V |

The wafer processing steps are as follows. After initial oxidation on the bare wafer, p+ strip implantation is performed using boron ion implantation. This is followed by deep n+, silicon oxide, polysilicon, metal, and passivation processing. Our fabrication plant does not have a silicon nitride facility to help remove pinholes in the oxide layer over the strips, so we rather optimize the photoresist material such that the fraction of strips with pinholes remains below 1%. The backside is implanted with n+ before applying a metal layer.

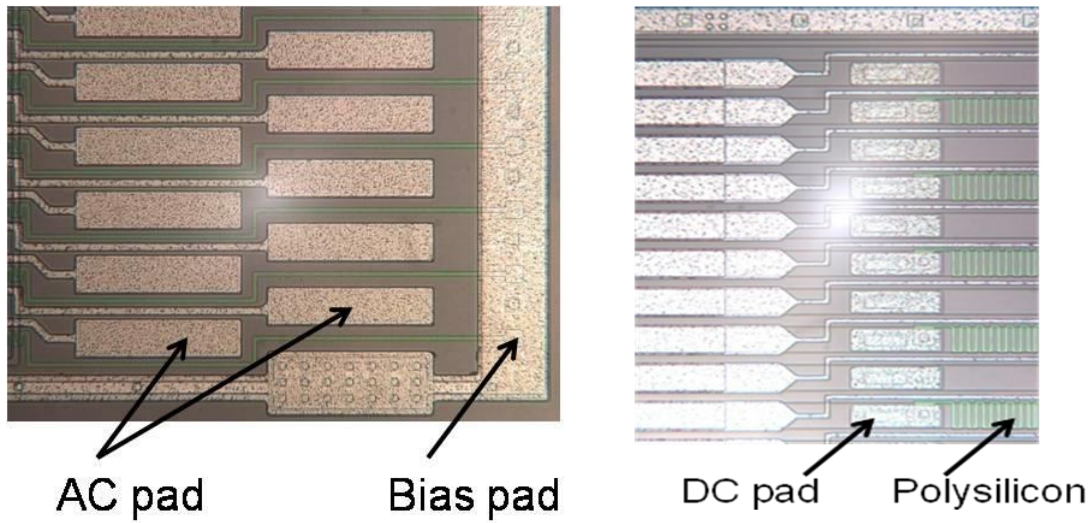

**Figure 2.** (Left) AC pad and bias pad, and (right) DC pad and polysilicon structure of the detector.

**2.3 Different batches of sensors**

In the second batch we developed single-sided sensors with resistivities of 10 to 20 kΩ-cm and a <100> crystal orientation. Typical total leakage currents were more than 10 µA at 100 V, which are very high compared to the required specification given in Table 1. In the third batch we used wafers with resistivities of 9 to 12 kΩ-cm and a <111> crystal orientation while keeping other specifications the same as the second batch. We were still unable to obtain much improvement in the leakage current. In the fourth batch we selected resistivity of 2 to 4 kΩ-cm and a crystal orientation of <111> with other specifications being the same as the second batch. We obtained very good results with five such detectors. The best one gave a total leakage current less than 1 µA at 300 V, while the other detectors were according to the required specifications of 1 µA at 100 V (See Section 4).

**3. Measurement setup**

We have a class 10,000 clean room with a manual Karl Suss PM8 probe station and a Keithley 237 source measure unit, Agilent 4284A LCR meters and a Keithley 6514 Electrometer for testing [4] the sensors. A decoupling box is used to connect the LCR meter and the source measuring unit in parallel to the sensor for CV measurements. We use a National Instruments PCI GPIB card installed in the computer, and all the instruments are connected via the GPIB cable to the PCI GPIB card. LABVIEW based programs were developed to control instruments used for carrying out various measurements with the silicon detectors. Figure 3 shows a schematic diagram of our measurement setup.

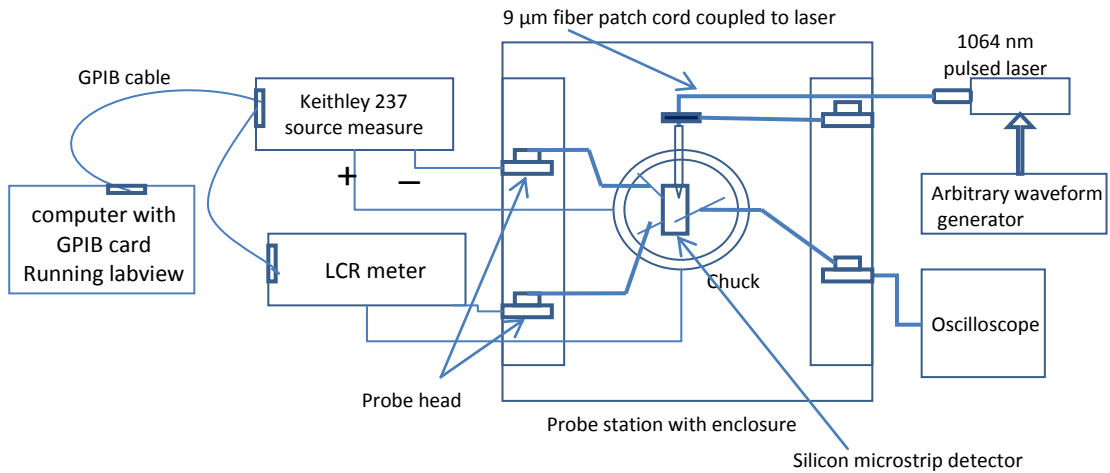

**Figure 3.** Schematic diagram of the measurement setup.

**4. Characterization**

I-V and C-V characterizations of the fourth batch sensors are presented in this section. We have also performed a laser test and a source measurement to obtain the charge collection efficiency. All measurements were performed at a temperature of 22°C with ±1°C tolerance.

**4.1 I–V test**

We applied a reverse voltage in steps of 10 V between the bias pad and the backplane, and measured the total leakage current in the sensor using the source measurement unit. Figure 4 shows the I–V characteristics of the fourth batch sensors. As evident from the plot, the total leakage current is less than 1 µA at 100 V, and the breakdown voltage ranges between 250 and 380 V. These values are according to our specifications.

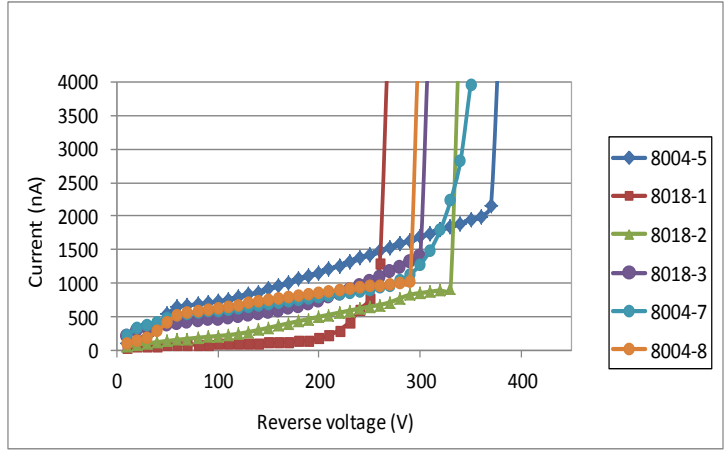

**Figure 4.** I–V characteristics of the fourth batch sensors.

### 4.2 C–V test

This test was aimed at determining the depletion voltage of the detector. We applied the reverse voltage in a similar way as for the I–V test (steps of 10 V), and measured the capacitance between the bias pad and the backplane at a frequency of 1 kHz. Figure 5 presents results of the bulk capacitance test for the fourth batch sensors. We found the depletion voltage to be around 70 V, the value beyond which the capacitance almost stabilizes. Furthermore, the capacitance value at depletion is in the range of 330-460 pF.

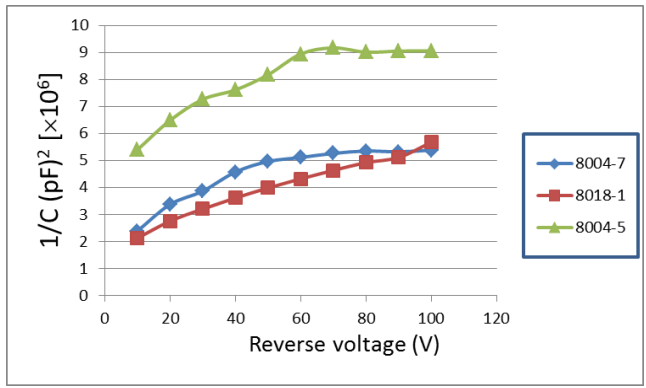

**Figure 5.** C–V characteristics of the fourth batch sensors.

In the fourth batch, the average polysilicon resistor value came out to be 3.75 MΩ. The fraction of strips with pinholes was less than 1%. The coupling capacitance measured at a frequency of 1 kHz was around 160 pF on each strip. All these values are consistent with the design specifications within tolerance.

### 4.3 Laser response of the sensors

To test the response [5] of the detectors we have used a Q-switched pulsed laser (Model QSL-1064-C12 from Lasermate [6]) with wavelength of 1064 nm and pulse duration of 10 ns. At this wavelength infrared light penetrates the whole 300 μm of silicon, thus measuring the entire bulk. We triggered the laser with an external TTL signal from a 100 MHz arbitrary waveform generator. The laser light was directed using a 9 μm patch fiber cable. The laser output energy was around 1.2 nJ, measured with an OPHIR energy meter. A pulse of 1.2 nJ energy translates to 7.5 GeV, which is equivalent to approximately 87,000 MIPs. We mounted the patch fiber cable on a probe head such that it can make fine X, Y and Z movements while keeping the cable close to the strips. With a reverse bias of 60 V applied to the sensor, the laser pulse was generated and the response measured on one of the AC pads. The signal from the AC pad was input to a linear fan-in/fan-out module. The inverted output signal was sent to a discriminator to generate a 100 ns gate signal for a CAEN V792N QDC [7]. The analog signal was delayed and given as input to the QDC module.

Figure 6 shows the measured QDC distribution having a peak charge of 7.76 pC. This measurement essentially provides a qualitative test of the sensors as one does not have a good control over the amount of light reflected at the surface or on the aluminum of the strip. Furthermore, there may have been some charge sharing with the neighboring strips that is not properly accounted for. So while we cannot precisely compare the measured charge with the expectation, the laser test still provides a useful check for the sensors in absence of a suitable low-noise amplifier. Particularly, we could see a clear difference between on- and off-strip signals to verify that the signal is indeed coming from the strip.

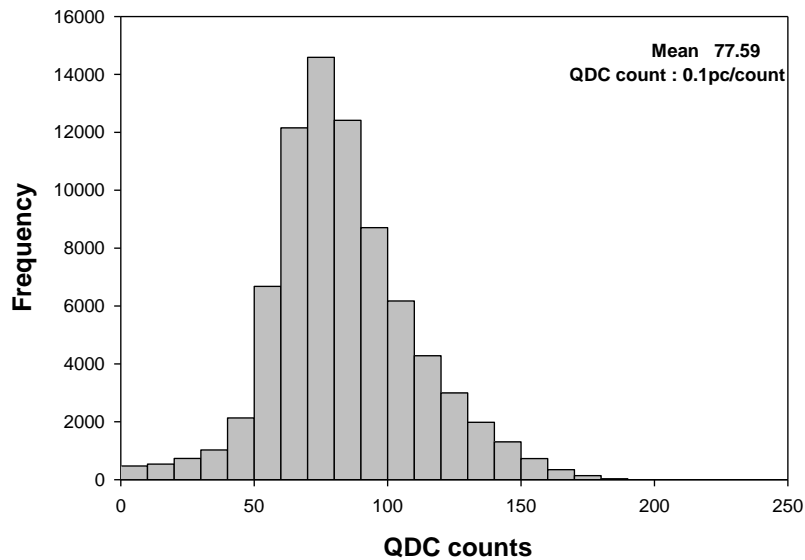

**Figure 6.** Measured QDC distribution.

### 4.4 Source measurement of the sensors

The response of one of the fourth-batch sensors (serial# 8004-8) to charged particles was tested with a $^{106}$Ru β source, emitting electrons of energy 3.54 MeV. The sensor was mounted on an aluminum plate and 128 strips were wire-bonded to an APV25 readout chip [8]. Figure 7 shows the picture of the assembly with a hybrid containing six chips and a glass pitch adaptor, necessary to match the pitch of the APV25 chip to that of the sensor.

The assembly was placed in the set-up and read out using a VME based data acquisition system at CERN. A scintillator coupled to a photomultiplier was used for the triggering. First a pedestal run was taken to extract the pedestal and the noise, $\sigma_{noise}$, defined as the RMS of the pedestal distribution. Then the source was placed over the sensor and the data taken again as a function of the bias voltage.

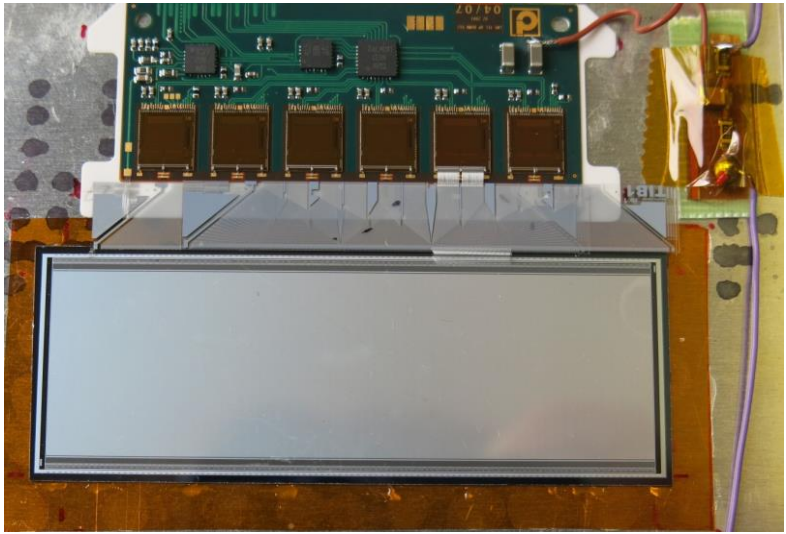

**Figure 7.** Sensor bonded to an APV25 chip, ready for testing.

The first step of the analysis is the subtraction of the pedestals from the raw data and the application of a correction for the correlated noise. It is followed by a search of a strip with a signal higher than a threshold of $3\times\sigma_{noise}$, the seed signal. Pulses recorded on the neighboring strips are added to the seed if they exceed a threshold of $2\times\sigma_{noise}$, and the sum is called a cluster signal.

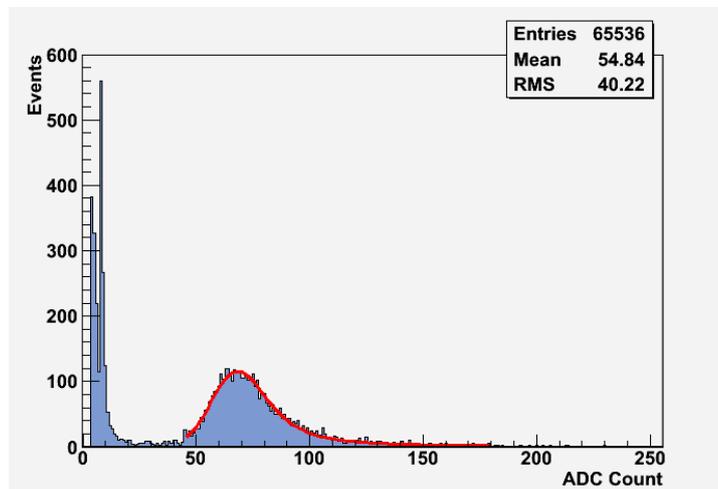

**Figure 8.** Cluster signal measured at 70 V with a fit (shown by the red curve), which is a convolution of a Landau distribution with a Gaussian function representing the noise.

Figure 8 shows the typical cluster signal measured at 70 V together with a fit, which is a convolution of a Landau distribution with a Gaussian function representing the noise. The absolute calibration of the signal to the number of electrons was done using an internal APV25 calibration pulse. Figure 9 shows the average noise distribution of 128 strips as a function of the bias voltage. One can see that the noise of about 650 e¯ is stable up to 110 V, the maximum voltage the sensor was tested to.

Figure 10 shows the signal as a function of the bias voltage. It reaches a plateau at about 70 V, which corresponds to the full depletion voltage measured on the sensor. The fitted most probable value of the signal at the plateau is about 23000 e¯, which gives a good signal to noise ratio of 35:1.

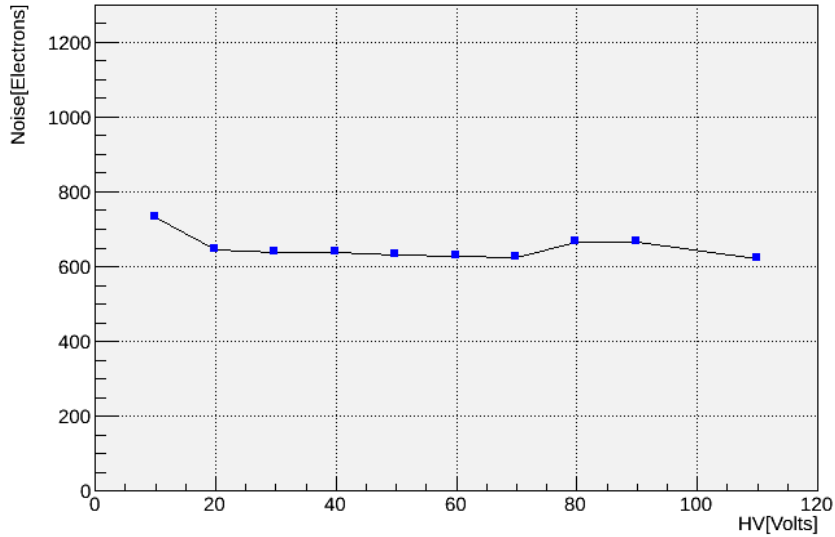

**Figure 9.** Average noise of 128 strips measured as a function of the bias voltage.

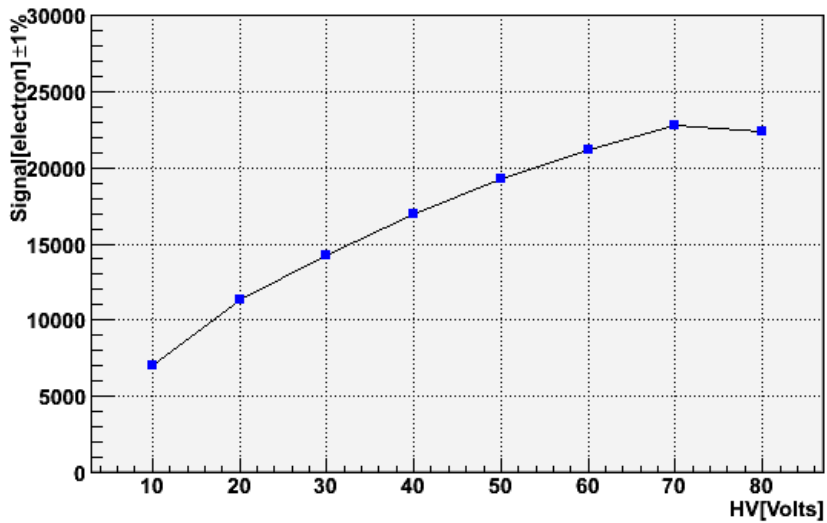

**Figure 10.** Signal generated by electrons from a $^{106}$Ru β source measured as a function of the bias voltage.

## 5. Conclusions and outlook

After fine-tuning the process parameters, we have successfully fabricated AC coupled single-sided silicon microstrip sensors for the first time in India. Various characterizations of the detectors were performed. In addition, their responses to a pulsed laser and a radioactive source were also studied. The sensors were found to be of good quality yielding an excellent signal-to-noise ratio. Looking to future, we are vigorously pursuing the R&D with double-sided microstrip sensors and sensors on a 6-inch wafer.

## Acknowledgments

We thank Alan Honma and Ian Mcgill (both from CERN) for preparing and bonding the sensor module. Helps from Alan, Anna Elliott-Peisert (CERN) and Markus Friedl (HEPHY) on reading the manuscript and providing useful feedbacks are warmly appreciated. The work is supported in parts by the Department of Atomic Energy, India.